\documentclass[twocolumn,aps,floatfix]{revtex4}
\usepackage{amssymb}
\usepackage{graphicx}
\usepackage{amsmath}

\begin{document}
\title{Resonant Einstein-de Haas effect in a rubidium condensate}

\author{Krzysztof Gawryluk,$\,^1$ Miros{\l}aw Brewczyk,$\,^1$ Kai Bongs,$\,^2$
        and Mariusz Gajda$\,^3$}

\affiliation{\mbox{$^1$ Instytut Fizyki Teoretycznej, Uniwersytet w Bia{\l}ymstoku,
                        ulica Lipowa 41, 15-424 Bia{\l}ystok, Poland}  \\
\mbox{$^2$ Institut f\"ur Laser-Physik, Universit\"at Hamburg, Luruper Chaussee
           149, 22761 Hamburg, Germany}  \\
\mbox{$^3$ Instytut Fizyki PAN, Aleja Lotnik\'ow 32/46, 02-668 Warsaw,
           Poland}  }

\date{\today}

\begin{abstract}
We numerically investigate a condensate of $^{87}$Rb atoms in an $F=1$ hyperfine state 
confined in an optical dipole trap. Assuming the magnetic moments of all atoms are initially 
aligned along the magnetic field we observe, after the field's direction is reversed, a 
transfer of atoms to other Zeeman states. Such transfer is allowed by the dipolar interaction 
which couples the spin and the orbital degrees of freedom. Therefore, the atoms in $m_F=0,-1$ 
states acquire an orbital angular momentum and start to circulate around the center of the 
trap. This is a realization of the Einstein-de Haas effect in systems of cold gases. We find
resonances which amplify this phenomenon making it observable even in very weak dipolar
systems. The resonances occur when the Zeeman energy on transfer of atoms to $m_F=0$ state
is fully converted to the rotational kinetic energy.

\end{abstract}

\maketitle

Magnetic effects in ultracold quantum gases have been subject to intense theoretical
and experimental studies during recent years. So far most of these investigations have
concentrated on short-range interactions as the dominant spin exchange process in spinor
condensates of $^{23}$Na~\cite{Stenger1998a} and
$^{87}$Rb~\cite{Matthews1998a,Schmaljohann2004a,Chang2004a}. These interactions result in
rich multi-component physics as demonstrated by the observation of phenomena like
magnetic phases~\cite{Stenger1998a,Schmaljohann2004a,Chang2004a}, coherent spin
dynamics~\cite{Schmaljohann2004a,Chang2004a,Chang2005a,Kronjaeger2005a}, domain
formation~\cite{Sadler2006a}, and a magnetically tuned spin mixing
resonance~\cite{resonance}. Including magnetic dipole-dipole interactions would further
enhance the richness of these systems and in particular their anisotropic nature is
expected to add completely new aspects. Already for relatively weak dipolar interactions
phenomena like the Einstein-de Haas effect~\cite{Ueda}, spontaneous
magnetization~\cite{Cheng2005a,Yi2006a}, squeezing and entanglement~\cite{Yi2006a} have
been predicted. Most of these studies concentrate on the recently achieved case of a
chromium Bose-Einstein condensate~\cite{Pfau}, as it is commonly believed that these
effects are practically unobservable in the widely available alkali condensates due to
the smallness of dipolar interactions in these systems. However, as pointed out
in~\cite{Yi2006a}, for $^{87}$Rb in the $F=1$ state the size of the dipolar interactions
as compared to the spin-mixing part of the short-range interactions reaches 10\%, such
that dipolar effects might be observable in this system.

In this Letter, we show that under the right conditions the dipolar interactions can
even dominate the dynamics of a $^{87}$Rb spinor condensate, making it a promising
candidate for the observation of the Einstein-de Haas effect \cite{Haas} as an important
milestone in dipolar quantum gas physics. In particular our calculations demonstrate the
existence of resonances in a spinor rubidium condensate that amplify the effect of dipolar
interactions. These resonances can be tuned by the magnetic field or by the trap geometry
and occur when the Zeeman energy fits the rotational kinetic energy per particle. The
resonances we find explore a new regime in comparison with that considered in Ref. \cite{Ueda}
for a $^{52}$Cr condensate, where the dipolar energy (not a kinetic one) is related to the
Zeeman energy.

In the second quantization notation, the Hamiltonian of the system we investigate is given by
\begin{eqnarray}
&&H = \int d^3r \left[ \hat{\psi}^{\dagger}_i(\mathbf{r}) \left(-\frac{\hbar^2}{2 m}
\nabla^2 + U_{ext}(\mathbf{r}) \right)   \hat{\psi}_i(\mathbf{r})
\right.  \nonumber  \\
&&\left. -\gamma \hat{\psi}^{\dagger}_i(\mathbf{r})\, \mathbf{B F}_{i,j}\,
\hat{\psi}_j(\mathbf{r}) + \frac{c_0}{2}\, \hat{\psi}^{\dagger}_j(\mathbf{r})
\hat{\psi}^{\dagger}_i(\mathbf{r})
\hat{\psi}_i(\mathbf{r}) \hat{\psi}_j(\mathbf{r}) \right.   \nonumber  \\
&&\left. +\frac{c_2}{2}\, \hat{\psi}^{\dagger}_k(\mathbf{r})
\hat{\psi}^{\dagger}_i(\mathbf{r})\,
\mathbf{F}_{ij} \mathbf{F}_{kl}\, \hat{\psi}_j(\mathbf{r}) \hat{\psi}_l(\mathbf{r})
\right]   \nonumber  \\
&&+ \frac{1}{2}\int d^3r\, d^3r' \hat{\psi}^{\dagger}_k(\mathbf{r})
\hat{\psi}^{\dagger}_i(\mathbf{r}') V^d_{ij,kl}(\mathbf{r}-\mathbf{r}')
\hat{\psi}_j(\mathbf{r}') \hat{\psi}_l(\mathbf{r})   \,,  \nonumber  \\
\label{Ham}
\end{eqnarray}
where repeated indices (each of them going through the values $+1$, $0$, and $-1$) are to
be summed over. The field operator $\hat{\psi}_i(\mathbf{r})$ annihilates an atom in the
hyperfine state $|F=1,i>$ at point $\mathbf{r}$. The first term in (\ref{Ham}) is the
single-particle Hamiltonian $(H_0)$ that consists of the kinetic energy part (with $m$ being
the mass of an atom) and the trapping potential $U_{ext}(\mathbf{r})$. The second term
describes the interaction with the magnetic field $\mathbf{B}$ with $\gamma$ being the
gyromagnetic coefficient which relates the effective magnetic moment with the hyperfine angular
momentum ($\boldsymbol{\mu}=\gamma\mathbf{F}$). The terms with coefficients $c_0$ and $c_2$
describe the spin-independent and spin-dependent parts of the contact interactions,
respectively -- $c_0$ and $c_2$ can be expressed with the help of the scattering lengths
$a_0$ and $a_2$ which determine the collision of atoms in a channel of total spin
$0$ and $2$. One has $c_0=4\pi\hbar^2(a_0+2a_2)/3m$ and $c_2=4\pi\hbar^2(a_2-a_0)/3m$
\cite{Ho}, where $a_0=5.387$\,nm and $a_2=5.313$\,nm \cite{a0a2}.
$\mathbf{F}$ are the spin-$1$ matrices. Finally, the last term describes the
magnetic dipolar interactions. The interaction energy of two magnetic dipole moments
$\boldsymbol{\mu}_1$ and $\boldsymbol{\mu}_2$ positioned at $\mathbf{r}$ and $\mathbf{r}'$
equals
\begin{eqnarray}
V_d = \frac{\boldsymbol{\mu}_1 \, \boldsymbol{\mu}_2}{|\mathbf{r}-\mathbf{r}'|^3}-3
\frac{
[\boldsymbol{\mu}_1 \, (\mathbf{r}-\mathbf{r}')] \,
[\boldsymbol{\mu}_2 \, (\mathbf{r}-\mathbf{r}')]}
{|\mathbf{r}-\mathbf{r}'|^5}
\label{mm}
\end{eqnarray}
and since $\boldsymbol{\mu}=\gamma\mathbf{F}$ one has
$V^d_{ij,kl}(\mathbf{r}-\mathbf{r}') = \gamma^2 \mathbf{F}_{ij} \mathbf{F}_{kl}
/ |\mathbf{r}-\mathbf{r}'|^3 - 3\gamma^2 [\mathbf{F}_{ij}  (\mathbf{r}-\mathbf{r}')]
[\mathbf{F}_{kl}  (\mathbf{r}-\mathbf{r}')] / |\mathbf{r}-\mathbf{r}'|^5 . $

The equation of motion reads
\begin{eqnarray}
i\hbar \frac{\partial}{\partial t}
\left(
\begin{array}{l}
\hat{\psi}_1 \\
\hat{\psi}_0 \\
\hat{\psi}_{-1}
\end{array}
\right)
=
({\cal{H}}_c + {\cal{H}}_d )
\left(
\begin{array}{l}
\hat{\psi}_1 \\
\hat{\psi}_0 \\
\hat{\psi}_{-1}
\end{array}
\right)   \,\,,
\label{Eqmot}
\end{eqnarray}
where the operator ${\cal{H}}_c$ originates from the single-particle and two-particle
but related to the contact interactions parts of the Hamiltonian (\ref{Ham}) whereas
${\cal{H}}_d$ corresponds to the dipole-dipole interactions. The diagonal part of
${\cal{H}}_c$ is given by
${\cal{H}}_{c 11} = H_0+(c_0+c_2)\, \hat{\psi}^{\dagger}_1 \hat{\psi}_1
+(c_0+c_2)\, \hat{\psi}^{\dagger}_0 \hat{\psi}_0
+ (c_0-c_2)\, \hat{\psi}^{\dagger}_{-1} \hat{\psi}_{-1}  , \;
{\cal{H}}_{c 00} = H_0+(c_0+c_2)\, \hat{\psi}^{\dagger}_1 \hat{\psi}_1
+c_0\, \hat{\psi}^{\dagger}_0 \hat{\psi}_0
+ (c_0+c_2)\, \hat{\psi}^{\dagger}_{-1} \hat{\psi}_{-1} , \;
{\cal{H}}_{c -1-1} = H_0+(c_0-c_2)\, \hat{\psi}^{\dagger}_1 \hat{\psi}_1
+(c_0+c_2)\, \hat{\psi}^{\dagger}_0 \hat{\psi}_0
+ (c_0+c_2)\, \hat{\psi}^{\dagger}_{-1} \hat{\psi}_{-1} $.
The off-diagonal terms that describe the collisions not preserving the projection of
spin of each atom (although the total spin projection is conserved) equal
${\cal{H}}_{c 10} = c_2\, \hat{\psi}^{\dagger}_{-1} \hat{\psi}_0  , \;
{\cal{H}}_{c 0-1} = c_2\, \hat{\psi}^{\dagger}_0 \hat{\psi}_1  $.
Moreover, ${\cal{H}}_{c 1-1} = 0$. On the other hand, for the ${\cal{H}}_d$ term
one has ${\cal{H}}_{d ij} = \int d^3r' \psi_n^{\dagger}(\mathbf{r}')V_{ij,nk}^d
\psi_k(\mathbf{r}') $. This term is responsible for the change of total spin projection
of colliding atoms. It turns out that when two atoms interact the total spin projection
$(\Delta M_F)$ can change at most by $2$. In particular, the diagonal elements of
${\cal{H}}_d$ lead to the processes with $\Delta M_F=\pm 1$. An example could be the
collision of two atoms in $m_F=1$ Zeeman state after which one of the atoms goes to the
component $m_F=0$. In addition to such processes, the off-diagonal terms of ${\cal{H}}_d$
introduce the interaction that change the spin projection $\Delta M_F$ by $\pm2$.
It happens when both atoms initially in the same state go simultaneously to the nearest
(in a sense of magnetic number $m_F$) state or in the case when atoms in different but
neighboring components transfer to the states shifted in number $m_F$ by $+1$ or $-1$.
There is no way for the atom being transfered directly from $m_F=1$ to the $m_F=-1$ state,
therefore, the populating of $m_F=-1$ component is a second order process.

Hence, the dipolar interaction does not conserve the projection of total spin of two
interacting atoms. Neither the projection of total orbital angular momentum is preserved
(see (\ref{com})). However, the dipolar interaction couples the spin and the orbital
motion of atoms as revealed by the last relation in (\ref{com})
\begin{eqnarray}
&&[V_d,F_{1z}+F_{2z}] \neq 0,\;\;\;\;   [V_d,L_{1z}+L_{2z}] \neq 0,       \nonumber  \\
&&[V_d,L_{1z}+L_{2z}+F_{1z}+F_{2z}]=0  \,.
\label{com}
\end{eqnarray}
Therefore, going to $m_F=0,-1$ states atoms acquire the orbital angular momentum and
start to circulate around the center of the trap. This is the realization of the famous
Einstein-de Haas effect in cold gases.

To solve the Eq. (\ref{Eqmot}) we neglect the quantum fluctuations and replace the field
operator $\hat{\psi}_i(\mathbf{r})$ by an order parameter $\psi_i(\mathbf{r})$ for each
component and apply the split-operator method. All integrals appearing in ${\cal{H}}_{d ij}$
are the convolutions and we use the Fourier transform technique to calculate them.
To find analytical formulas for the Fourier transforms of the components of the
convolutions that do not change during the evolution we apply the regularization procedure
described in Ref. \cite{Goral}.

The gyromagnetic coefficient for $^{87}$Rb atoms in an $F=1$ hyperfine state is positive
and equals $\gamma=\frac{1}{2} \mu_B /\hbar$. We prepare an initial state of the condensate
as the one with all magnetic moments aligned along the magnetic field, i.e., all atoms are
in $m_F=1$ component. To this end, we run the mean-field version of Eq. (\ref{Eqmot}) in
imaginary time while the magnetic field is turned on (and equal to $B=0.73$\,mG
for a spherically symmetric trap with the frequency $\omega=2\pi \times 100$\,Hz). Then we
reverse the direction of the magnetic field and look for the transfer of atoms to other
Zeeman states.

\begin{figure}[thb] \resizebox{3.4in}{2.5in}
{\includegraphics{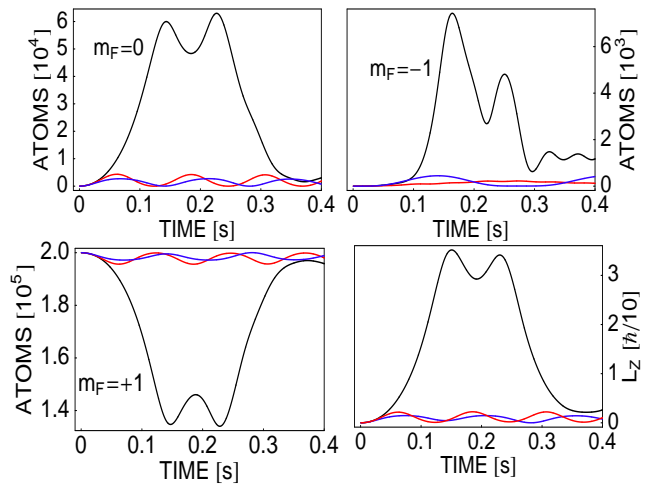}}
\caption{(color online). Transfer of atoms to $m_F=0,-1$ Zeeman states as a function of
time. Initially, $N=2\times 10^5$ atoms were prepared in $m_F=1$ component in a spherically
symmetric trap with the frequency $\omega=2\pi \times 100$\,Hz. The residual magnetic field
equals $-0.029$\,mG (black lines -- on-resonance case) and $B=-0.015$\,mG, $-0.036$\,mG 
(blue and red lines, respectively -- off-resonance case). The lower panel shows:
the number of atoms in $m_F=1$ state and the time dependence of the orbital angular 
momentum per atom.}
\label{transfer}
\end{figure}

\begin{figure}[thb] \resizebox{2.4in}{2.8in}
{\includegraphics{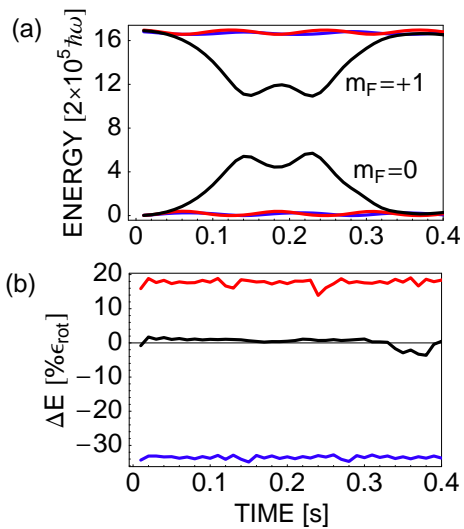}}
\caption{(color online). Total energy for $m_F=1,0$ components as a function of time (upper
frame). All parameters are the same as in Fig. \ref{transfer}. The resonance happens for 
$B=-0.029$\,mG (black lines) whereas the off-resonance cases are represented by
$B=-0.015$\,mG (blue lines) and $B=-0.036$\,mG (red lines). The lower frame shows 
$\Delta E=(E_{kin}^{+1}/N_{+1}+\mu B - \epsilon_{rot})/\epsilon_{rot}$, where
$\epsilon_{rot}=E_{rot}/N_{0}$. }
\label{epa}
\end{figure}

Our starting condition (all atoms in $m_F=1$ state) suppresses the short-range spin
dynamics and initially the $m_F=1$ state is depleted only due to the dipolar interaction.
Usually, we observe a small number of atoms going from the Zeeman state $m_F=1$ to the
$m_F=0,-1$ states. However, on resonance (see Fig. \ref{transfer}) the transfer to the
other states can be of the order of the initial population of $m_F=1$ component. This
transfer is as large as in the case of chromium condensate \cite{Ueda} despite the fact
that the dipolar energy ($\mu^2 n$, where $n$ is the atomic density) is approximately
$100$ times smaller. The only difference is that the time scale corresponding to the
maximal transfer is about $100$\,ms, i.e., $100$ times longer than for chromium. This can
be understood as follows. For $^{87}$Rb the dipolar energy is the smallest energy of the
problem and is only a perturbation as compared to the kinetic rotational energy and
the Zeeman energy. An efficient population transfer from $m_F=1$ to $m_F=0$ due to the
dipolar interactions is only possible, when the total energies (the sum of mean-field, trap,
kinetic, and Zeeman energies) in these states are approximately equal. Fig. \ref{epa}a 
clearly shows that the total energy tends to equalize on resonance, which is not true in
the off-resonant case. Numerics also shows (Fig. \ref{epa}b) that the above condition can 
be fulfilled only then when the Zeeman energy fits the rotational kinetic energy: 
$\mu B_{res} = \epsilon_{rot}$. It means that the resonant magnetic field is inversely 
proportional to the magnetic moment of an atom
\begin{eqnarray}
B_{res} \propto 1/\mu    \,.
\label{rescon}
\end{eqnarray}
Surprisingly, the smaller atomic magnetic moment the larger value of the resonant magnetic
field. For chromium condensate, however, the dipolar energy is larger
than the kinetic energy. Therefore, in this case the resonance condition should be derived
by relating the Zeeman and the dipolar (not the kinetic one) energies, i.e.,
$\mu B_{res} = \mu^2 n$ \cite{Ueda}. This results in a condition $B_{res} \propto \mu$ which
differs qualitatively from (\ref{rescon}).

The maximal transfer is reached at time which is of the order of characteristic
time scale determined by the dipolar interactions $(\hbar/\mu^2 n)$. Since the magnetic
moment of $^{87}$Rb atom is $12$ times smaller than that of $^{52}$Cr we have to wait
hundreds of milliseconds (not a fraction of millisecond as in Ref. \cite{Ueda}) to see
the action of resonance.

Fig. \ref{epa} illustrates the ideas just discussed. In the upper frame the total
energies of $m_F=1,0$ components are plotted as a function of time both in on- and 
off-resonance cases showing that the resonances we find is a dynamical phenomenon.   
The transfer gets maximal when the energies approach each other (perhaps crossing both
curves would require the dynamical tuning of the resonance by changing the magnetic field). 
Contrary, almost no transfer of atoms occurs when the energy curves keep away. Simultaneously,
the lower frame proves that on resonance the Zeeman energy (in fact, together with
the kinetic energy) is transfered to the rotational kinetic energy of atoms in $m_F=0$ 
component.

\begin{figure}[thb] \resizebox{2.in}{2.in}
{\includegraphics{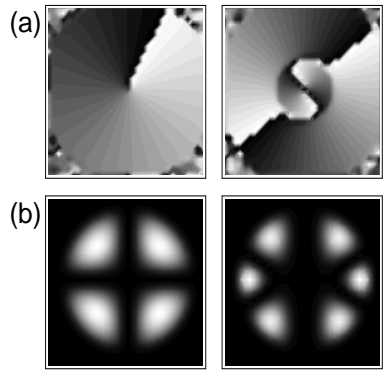}}
\caption{(a) Phase in the 'xy' plane of $m_F=0$ (left frame) and $m_F=-1$ (right frame)
spin components.
(b) Density in the 'xz' plane (the 'z' axis goes vertically). All parameters are as in
Fig. \ref{transfer} (on-resonance case) and the density cuts are taken at $140$\,ms.}
\label{cuts}
\end{figure}
Huge transfer of atoms to $m_F=0,-1$ states is the realization of the
Einstein-de Haas effect in cold gases. Fig. \ref{cuts}a proves that the vortices are
generated in $m_F=0,-1$ components (though already the lower-right frame in 
Fig. \ref{transfer} suggests that atoms in states $m_F=0,-1$ rotate around the quantization 
axis). For $m_F=0$ component the phase of the order parameter winds up by $2\pi$ which means 
that a singly quantized vortex is formed in this state. At the same time in the $m_F=-1$
component a doubly quantized vortex is formed (the phase winds up by $4\pi$) as a result
of total angular momentum conservation. Fig. \ref{cuts}b shows the typical density
patterns in the $m_F=0,-1$ states. The density is fragmented and the number of rings in
the $m_F=0,-1$ components results from the symmetry of dipolar interaction. For example,
since the dipolar interaction transforms as a spherical tensor of rank-2 in both spatial
and spin spaces, it spatially acts as the spherical harmonic $Y_{21}$ when atoms go from
$m_F=1$ to $m_F=0$ state. And, since the initial state is spherically symmetric ($\propto
Y_{00}$) it induces the spatial behavior $\propto Y_{21}$ in $m_F=0$ component.
Therefore, the density in $m_F=0$ state shows two rings. Similar fragmentation was
already predicted in the case of $^{52}$Cr condensate in Ref. \cite{Ueda}.

\begin{figure}[bth]
\resizebox{2.4in}{2.8in}
{\includegraphics{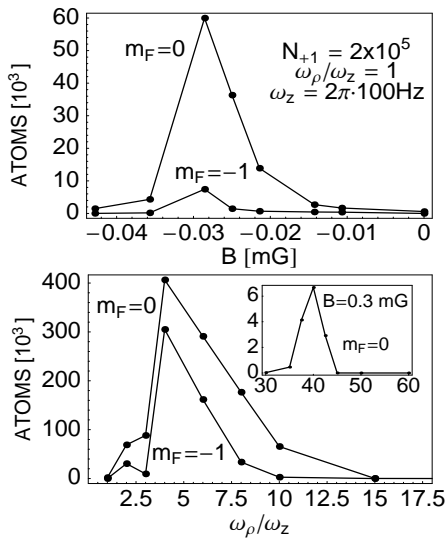}}
\caption{Maximal transfer of atoms to $m_F=0,-1$ states as a function of the residual
magnetic field (upper frame) and the trap geometry (lower frame). For lower frame
$\omega_z=2\pi \times 100$\,Hz, $B=-0.073$\,mG, and $N_{+1}=8\times 10^5$.
Inset shows the resonance at $B=-0.3$\,mG for $\omega_z=2\pi \times 20$\,Hz and
$N_{+1}=10^5$.}
\label{res}
\end{figure}

\begin{figure}[thb] \resizebox{3.0in}{1.8in}
{\includegraphics{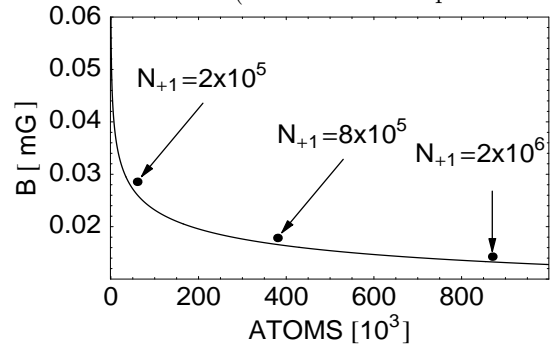}}
\caption{Comparison between numerics (balls) and the Thomas-Fermi approximation. Solid line
indicates the value of the magnetic field at resonance (for spherically symmetric trap) as
a function of number of atoms in $m_F=0$ state. }
\label{TF}
\end{figure}

Fig. \ref{res} (upper frame) shows the position and the width of the resonance displayed
in Fig \ref{transfer}. Similar behavior is observed when the value of the reversed magnetic
field is kept constant and the trap geometry is changed (lower frame). Here,
the maximal transfer of atoms is obtained in a cigar trap with the aspect ratio
$\omega_{\rho}/\omega_z=4$ with almost $50\%$ efficiency at $B=-0.073$\,mG. The inset shows
the resonance at experimentally easier to control value of magnetic field $B=0.3$\,mG but
still detectable number of atoms in $m_F=0$ state.
To understand quantitatively the resonance we start from the condition discussed earlier:
$\mu B_{res} = E_{rot}/N_{0}$, where $E_{rot}$ is the rotational energy of
$m_F=0$ component which is assumed to be a singly quantized vortex, given within the
Thomas-Fermi approximation by
$\psi_0(\rho,\phi,z)=[(\lambda-m\omega^2(\rho^2+z^2)/2-\hbar^2/(2m\rho^2))/c_0]^{1/2} \times
e^{i \phi}$. Here, the chemical potential $\lambda$ is obtained by the requirement that the
number of atoms in $\psi_0$ state equals $N_0$. One can tune to the
resonance in two ways: a) by adjusting the magnetic field $B$; b) by changing the trap geometry
what influences the rotational energy entering resonance condition and keeping the magnetic
field constant. The resulting curve for the spherically symmetric trap is plotted in
Fig. \ref{TF}. The numerical results are marked by balls and correspond to the initial number
of atoms in $m_F=1$ component equal to $2\times 10^5$, $8\times 10^5$, and $2\times 10^6$
showing a good agreement. For other systems, e.g. $^{52}$Cr, the condition
$\mu B_{res} = E_{rot}/N_{m_F}$ suggests that the value of the resonant magnetic field is
even $\approx 10$ times smaller since $\mu_{Cr}/\mu_{Rb}=12$ and $E_{rot}/N_{-2}$
for chromium looks similar as $E_{rot}/N_{0}$ in the rubidium case.

In conclusion, we have shown the existence of dipolar resonances in rubidium spinor
condensates. The resonances occur when the Zeeman energy of atoms in
$m_F=1$ component, while transferring to $m_F=0$ state, is fully converted to the rotational
kinetic energy. This is so far an unexplored regime. Symmetries of the dipolar interaction
force the atoms in $m_F=0,-1$ states to circulate around the quantization axis and form singly
and doubly quantized vortices, respectively. Therefore, dipolar resonances is a route to the
observation of the Einstein-de Haas effect (as well as other phenomena related to the dipolar
interaction) in weak dipolar systems.

\acknowledgments We are grateful to J. Kronj\"ager, J. Mostowski, and K. Rz\c a\.zewski
for helpful discussions. M.B. and M.G. acknowledge support by the Polish KBN Grant
No. 1 P03B 051 30. K.G. thanks the Polish Ministry of Scientific Research Grant Quantum
Information and Quantum Engineering No. PBZ-MIN-008/P03/2003.

\end{document}